\documentstyle[aps,prl,epsfig,twocolumn]{revtex}
\title{ A new Manifestation of Atomic Parity Violation in Cesium:\\
a Chiral Optical Gain induced by linearly polarized 6S-7S Excitation }
  \author{ J. Gu\'ena, D. Chauvat$^a$, Ph. Jacquier, E. Jahier, M. Lintz, S.
Sanguinetti$^b$, A. Wasan, M.A.~Bouchiat}
\address{ Laboratoire Kastler Brossel , 24 Rue Lhomond, F-75231 
Paris  Cedex 05, France \\
  (\today)}
\author{ \vspace{0.5mm} A.V.~Papoyan, D. Sarkisyan}
\address{Institute for Physical Researches, Ashtarak-2, 378410, Armenia}

\author{\parbox{410pt}{ \vglue 0.3cm \footnotesize\sf
  We have detected, by using stimulated emission, an Atomic Parity
Violation (APV) in the form of a chiral optical gain of a cesium 
vapor on the 7S - 6P$_{3/2}$ transition,
consecutive to linearly polarized 6S-7S excitation.  We demonstrate 
the validity of this detection method of APV, by presenting a
  9$\%$ accurate measurement of expected sign and magnitude. We 
underline several
advantages of this entirely new approach in which the cylindrical 
symmetry of the set-up can be fully exploited. Future measurements at
the percent level will provide an important cross-check of an 
existing more precise result obtained by a different method. ~~~~
  }\vspace{-1.5cm}}

\begin{document}
\maketitle
  APV experiments are important because they probe the electron-hadron 
$Z^0$ exchange in conditions totally
different from those encountered in high energy experiments 
\cite{bou97}. The Boulder group \cite{woo97,ben99} has made quite an
achievement by performing  a calibrated \cite{bou881} measurement of 
the parity violating  6S-7S
transition amplitude ${\rm E}_1^{pv}$,  at a level of precision 
better than one per cent. The importance of this work is
illustrated by the abundant literature triggered by the result.  An 
atomic physics calculation is required to
arrive at a precise determination of the weak nuclear
charge $Q_W$\cite{bou97}, measuring the interaction strength. Atomic
theorists  have met the real challenge of reducing their calculation 
uncertainty below the existing value of one per cent
\cite{blu90}. Subtle Breit \cite{der01} and radiative \cite{mil01} corrections,
previously omitted, were found to contribute at the few tenths
  of a percent level. Several authors \cite{ben99,fla02} have 
announced a deviation of 2.5 or 2.2$\sigma$ between the
empirical determination of
$Q_W$ and the theoretical value expected in the Standard Model. 
Therefore, particle physicists have searched for the
possible implications of such a deviation
when taken seriously.
The best way to
account for it, without causing conflict with high energy results, is 
to admit the existence of an additional
neutral vector boson $Z_0'$, whose mass differs according to the 
models, but always falls around a fraction of
TeV~\cite{cas99}.

In view of such an important implication,
an independent measurement obtained by a totally different method, 
offering a cross-check of the result, would be extremely
valuable.  Our experiment has been designed to achieve this. Our 
approach differs from the Boulder's one in two respects: a
different emphasis given to signal-to-noise ratio (SNR) versus 
background level, and the choice of a new APV physical
observable.  The idea arose after our first experiment~\cite{bou82} 
succeeded in illustrating the interest of measurements in
the forbidden 6S-7S Cs transition. It became clear that a novel 
detection scheme was required to improve their sensitivity. The
polarization analysis of the fluorescence light
  collected only one thousandth of the photons emitted by the excited 
7S atoms. We proposed a more efficient
scheme based on stimulated emission from 7S to 
6P$_{3/2}$\cite{bou85}. As shown in Fig.1, an intense pulsed laser 
excites the 6S - 7S
transition in a time short compared to the 7S lifetime, in a {\it 
longitudinal electric field} $\vec E_l$, giving rise to a large 
population
inversion and {\it transient amplification of a probe beam} tuned at 
resonance for the 7S - 6P$_{3/2}$ transition. Thus, nearly all atoms
contribute to the signal. The dependence of the amplification on the 
relative orientation of the linear excitation polarization $\hat
\epsilon_{ex}$ and the linear probe one
$\hat \epsilon_{pr}$, is the effect which demonstrates APV. {\it In 
two mirror-image configurations the excited vapor exhibits different
gains, depending on whether the trihedron formed by $\vec E_l, \hat
\epsilon_{ex}$ and $\hat \epsilon_{pr}$, is either right or left}. In 
the present paper, we report on the
first APV manifestation via such {\it a chiral optical gain} which 
validates our approach.

The pseudo-scalar  $ (\hat \epsilon_{ex} \cdot \hat \epsilon_{pr})( 
\hat \epsilon_{ex} \wedge \hat \epsilon_{pr}\cdot \vec E_l)$
appearing in the gain leads to
{\it a tilt  $\theta^{pv}$ of its optical axes with respect to the 
symmetry planes} of the experiment (Fig.1).
This tilt is given by the ratio of the PV  E$_1$ amplitude to the 
vector part of the Stark induced amplitude~\cite{bou97},
   $\theta^{pv} = - \rm{Im E}_1^{pv}/\beta {\it E_l} $, about one
$\mu$rd in current experimental conditions.  We note that it is odd 
under  $\vec E_l$ reversal.
  In our experiment,  two mirror-image configurations, associated with 
the beams  coming out
  from the polarizing beam splitter (PBS), are observed 
simultaneously, as shown in Fig.1.  At the cell input $\hat 
\epsilon_{pr}$
is aligned along one of the optical axes in absence of APV, either 
parallel ($\parallel $) or perpendicular ($\perp $) to  $\hat 
\epsilon_{ex}$.
During propagation through the vapor, APV  induces  a small tilt of 
$\hat \epsilon_{pr}$
towards the direction of largest gain. This linear dichroism should 
not be confused with a true optical rotation!
It gives  rise to an imbalance, odd under $\vec E_l$ reversal,  in
our two channel polarization analyzer, operating in balanced mode  in 
absence of APV. In an ideal configuration, the imbalance between
the optical signals in the two channels, $S_1$ and $S_2$, yields a 
measurement of the left-right (L-R) asymmetry\cite{bou96}:
\begin{equation}
A_{LR}=\frac{(S_1 - S_2)}{ (S_1 + S_2)} = -2  \theta^{pv} \big[ \exp 
{ (\eta {\cal A}) } -1\big]\, .
\end{equation}
  \noindent Here ${\cal A} $ is the optical density for the probe and 
$\eta$ the gain anisotropy, equal to $
(\alpha_{\perp}-\alpha_{\parallel})/2 \alpha_{\parallel}$, where 
$\alpha_{\parallel} $ (resp. $\alpha_{\perp})$ is the
probe gain per unit length for $\hat \epsilon_{ex} \parallel \hat 
\epsilon_{pr}$ (resp.  $\hat \epsilon_{ex} \perp \hat \epsilon_{pr})$.
In these APV measurements, we selected the hyperfine component of the 
probe transition, the $7S_{1/2, F=4} - 6P_{3/2,
F=4}$ line,  leading to the largest value of $\eta$, hence the 
largest L-R asymmetry.  As  Eq. 1 shows, the greater the optical
density, the larger the amplification of both the  probe beam
and the L-R asymmetry. We took advantage of this by exploiting the 
high optical densities available in a vapor cell, and by increasing 
the
excitation energy and the $\vec E_l$ field magnitude. The exponential 
growth of the asymmetry with the applied field is the attribute of
detection by stimulated emission, contrasting with fluorescence 
detection where the asymmetry
$\propto  \rm{Im\,E_1^{pv}}/
\beta {\it E_l} $ decreases when $E_l$ is increased. It should be 
stressed that a precise calibration procedure, independent on 
line-shape,
optical density and saturation, has been devised and carefully tested
\cite{gue97}. This consists in deliberately tilting the direction of 
$\hat \epsilon_{ex}$ with respect to $\hat \epsilon_{pr}$ by
a small precisely calibrated angle $\theta_{cal}$. It can be shown 
that this causes the probe polarization to rotate by an angle
$k \theta_{cal}$, {\it with the same proportionality factor k} as the 
APV angle $\theta^{pv}$. On the other hand, due to the
absence of magnetic fields, all atoms belonging to a given hyperfine 
state contribute and line shape problems coming from
overlapping Zeeman components are avoided. In addition, the suppression
  in the longitudinal $E_l$-field configuration of the
$M_1$-Stark interference effect, which is a potential source of 
systematic effect,  is welcome.
A further attractive feature is the cylindrical symmetry exhibited by 
the experiment: the signal is expected to remain invariant
under simultaneous rotations of  the polarizations $\hat 
\epsilon_{ex}$ and $\hat \epsilon_{pr}$ about the common beam 
direction.
This feature has enabled us to discriminate against possible 
systematics arising from stray fields.

Despite  the fact that our scheme  combines several attractive new 
features, the
observation of the APV chiral optical gain  remains an experimental 
challenge.  To make
the method work we have had to solve often unexplored  experimental 
problems. Besides
a  high precision differential polarimeter operating in pulsed regime 
\cite{gue97}, our experiment requires both the pulsed excitation laser
and the probe laser systems  to satisfy uncommon specifications 
described elsewhere\cite{gue98}.
  For instance, it is crucial to restrict  the detection to the  short 
time interval ($\sim$ 20 ns) during which the vapor  acts as an
amplifier,  with the help of  a fast optical switch on the probe 
laser beam.  Another  major  problem was
actually to generate a pulsed longitudinal uniform  $\vec E_l$-field of
$\simeq$ 2 kV/cm inside a Cs vapor cell at the useful atomic
densities ($\simeq 10^{14}$/cm$^{-3}$) and over a length of 80 mm.
Moreover, the initial glass cells \cite{gue98} had to be replaced by 
alumina ones consisting of an alumina tube,
  closed by sapphire windows glued at both ends\cite{sar89}.
The  surface electrical conductivity of alumina being much lower than 
that of glass
\cite{bou99}, stray magnetic fields induced by electric surface 
currents are thereby suppressed.
With alumina, one can use external electrodes if need be 
\cite{jah01}. The sapphire windows, unlike glass ones, retain
excellent transparency under intense laser illumination  in presence 
of the Cs vapor. Finally, one can heat alumina cells to
250-300$^{\circ}$~C  to thermodissociate troublesome Cs$_2$ dimers. 
As a	 consequence, the $\vec E$-field map
distorsions  coming from  dimer photoionization by  the excitation 
laser beam \cite{bou92} are strongly reduced.
One remaining problem, however,
arises from  the photoelectrons emitted  by the windows. We have 
shown that these are considerably  amplified by secondary
electron emission (SEE) from  grazing  collisions with the walls.
  An efficient  solution has consisted in machining annular grooves with
sharp tips in the alumina walls (1~mm apart), making grazing 
incidence unlikely \cite{gue02}.
This modification of the internal surface
has led to strong reduction of the $\vec E, \vec B$ stray fields and enabled us
  to perform the APV
measurements reported here.
\begin{figure}
\centerline{\epsfxsize=87mm \epsfbox{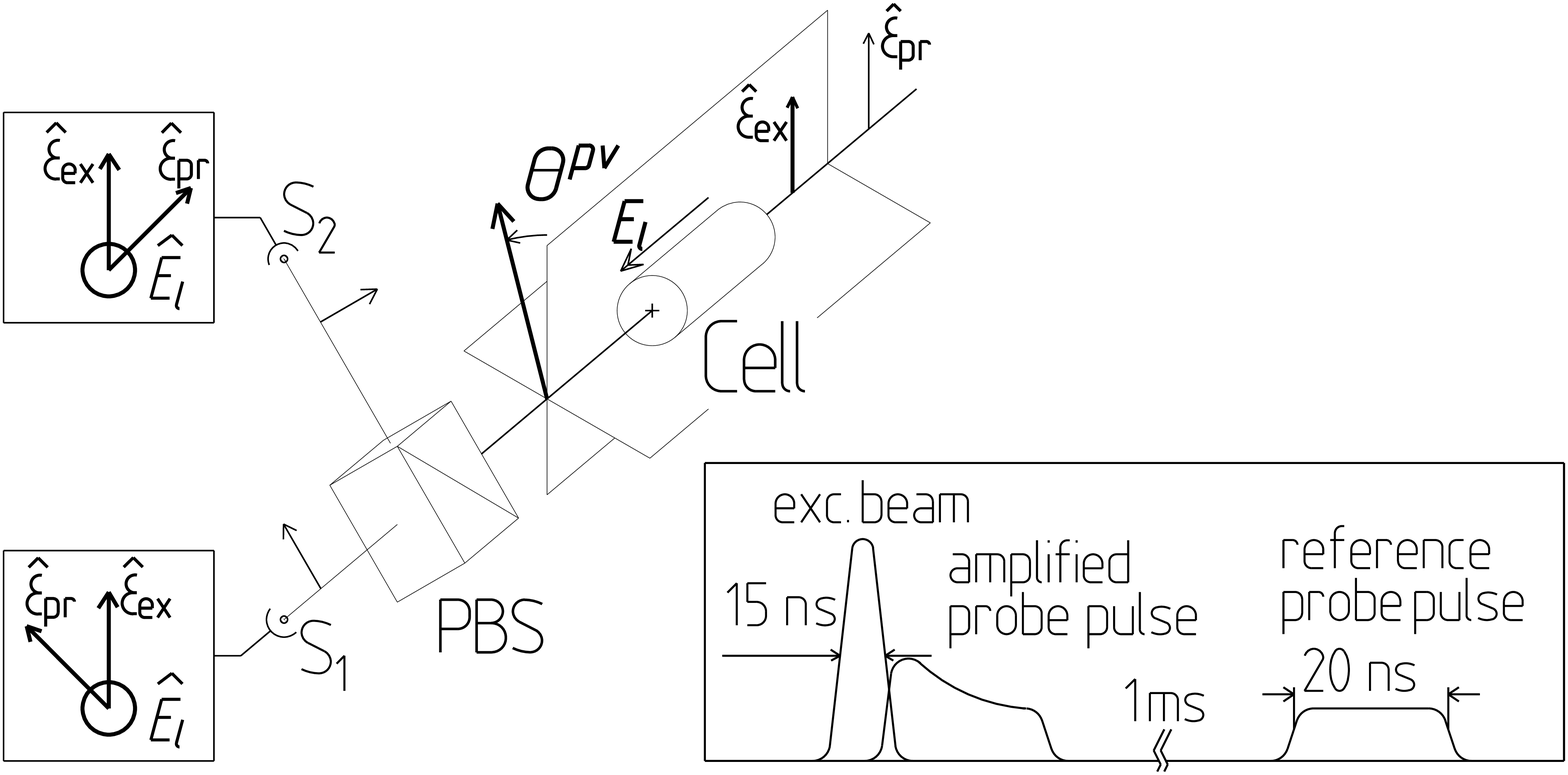}}
\vspace{3mm}
\caption {\footnotesize Schematic of the experiment showing the two 
orthogonal symmetry planes defined
by the electric field $\vec E_l$ and
the linear excitation polarization  $\hat
\epsilon_{ex}$. APV gives rise to a tilt
$\theta^{pv}$  of the optical axes of the excited vapor out of those 
planes. The incoming probe polarization $\hat \epsilon_{pr}$ provides 
a
superposition of the right and left-handed $\hat \epsilon_{ex}, \hat 
\epsilon_{pr}, \vec E_l$ configurations
analyzed. The probe amplification difference is
directly extracted from the optical signals S$_1$, S$_2$, recorded in 
each channel of the Polarizing Beam Splitter
(PBS). Inset: timing of the experiment repeated at 100 Hz.}
\end{figure}

   The signature of the PV signal first relies on a doubly 
differential feature of the
polarimeter output. The imbalance at each amplified probe pulse, 
$D^{amp}=\frac{(S_1-S_2)}{(S_1+S_2)}$
is compared to that of four consecutive probe pulses, $D^{ref}$, 
measured when all 7S
excited atoms have decayed. Thus  the doubly differential signal, 
$D^{amp}-D^{ref}$, selects at the laser repetition rate
($\sim$ 100 s$^{-1}$) a {\it true atomic effect}, $D_{at}$, free from 
the probe polarization rotation (independent of the excited
atoms) arising from polarization defects present on the probe beam path.
Our measurement involves four different physical parameter
reversals, listed in order of increasing period. 1) reversal of
$\theta_{cal}$, every 0.25 s, providing for real time calibration; 2) 
reversal of the $\vec E_l$ field
every 0.5 s (with 60 ms dead time); 3) switching of the half-wave 
plate $ (\frac{\lambda}{2})^{det}$ in front of the polarimeter, which
performs a symmetry of the outgoing probe polarization with respect 
to the symmetry plane of the experiment, every 2~s: it allows us to
discriminate between true polarization rotations and instrumental 
imbalances \cite{gue97}; 4) switching of another half-wave plate $
(\frac{\lambda}{2})^{pr}$ which acts on the input probe polarization, 
from para ($\hat \epsilon_{ex} \parallel \hat
\epsilon_{pr}$) to ortho ($\hat \epsilon_{ex} \perp \hat
\epsilon_{pr}$)  configuration
for discrimination between linear dichroism and optical
rotation \cite{gue97}.
Since performing twice a given reversal leaves the
system unchanged, for each reversal we introduce one binary variable: 
$\sigma_{cal}, \sigma_{_E}, \sigma_{det},
\sigma_{pr} = \pm1$.
Thus the complete signature of the APV
signal involves an average over all the $2^4$ possible states. For a 
given $f(\sigma$), we define the average, $< f(\sigma)>_{\sigma} =
\frac{1}{2}(f(1) + f(-1))$, i.e. the $\sigma$-{even} part. This 
implies that  $<  \sigma f(\sigma)>_{\sigma} = \frac{1}{2}(f(1) -
f(-1))$ yields the $\sigma$-{odd} part. The determination of the PV 
calibrated linear
dichroism involves the construction of the following quantity:
\begin{eqnarray}
G = \theta_{cal}  \left <    \hspace{-1mm} \sigma_{_E} 
\hspace{-1.5mm}  \left  [\frac{ < \sigma_{det} D_{at}
(\lbrace \sigma_j  \rbrace) >_{\sigma_{det} \sigma_{cal}}}{< 
\hspace{-1mm} \sigma_{det} \sigma_{cal}D_{at}(\lbrace \sigma_j 
\rbrace )
\hspace{-1mm}> _{\sigma_{det}
\sigma_{cal}}}
\right ]   \right >_{\hspace{-1mm}\sigma_{_E} \sigma_{pr}}
\end{eqnarray}
\noindent We stress that an imperfect $\vec E_l$ reversal leading to 
a $\vec E_{l,odd}$ contribution to ${\cal A}$ does not
affect the RHS of Eq.~2.

Moreover, measurements with different orientations of $\hat 
\epsilon_{ex}$ provide important tests for
systematic effects associated with stray transverse $\vec E_T$ and 
$\vec B_T$ fields (or beam misalignment), since they are
related to two preferred directions in the transverse plane. In fact, 
four orientations suffice  because of the structure of the
Stark induced  dipole. A set of four supplementary $\lambda /2$ 
plates (two on each beam) enables us to perform successive rotations 
of
$\hat \epsilon_{ex}$ (and $\hat \epsilon_{pr}$) by steps of 
45$^\circ$, (directions $\hat i = \hat x, \hat u, \hat y, \hat v $), 
while keeping
the analyzer fixed \cite{gue98}.    From the four measurements, $G_i $, we
extract two values of the isotropic part, $ \frac{1}{2}(G_x+G_y)=
\frac{1}{2}(G_u+G_v)$, expected - and actually found - to be equal to 
within the noise and to average to
$\theta^{pv}$. The presence of an anisotropic part manifests itself 
through the differences  $ D_{xy}=\frac{1}{2}(G_x-G_y)$ and $
D_{uv} = \frac{1}{2}(G_u-G_v)$. The isotropy test consists in plotting one
point in the cartesian coordinate system $D_{xy} , D_{uv}$, for each 
cycle of four orientations  of $\hat \epsilon_{ex}$  and to
look for a possible deviation of the center of gravity of the cloud 
of points associated to all data. Clear anisotropies were sometimes
observed. However, since their correlation with the isotropic part is 
not significant, the latter is negligibly affected. The two isotropic
values obtained for each PV cycle (every
$\sim$ 8 mn) were averaged together over a number N of
cycles.

Finally, for practical reasons, we are obliged to carry out a further 
reversal. Although the ideal configuration requires perfect alignment
of both laser beams along the cell axis, we have to concede a small 
tilt angle of the cell axis ($\psi\sim$ 3 mrd), in order to
avoid an excess of noise in the polarimeter. Since such a tilt breaks 
the symmetry, we reverse its sign every few hours and we take the
average of the results obtained with both tilts, affected in practice 
by similar statistical noise:
$\theta^{pv}_{exp}= \frac{1}{4} < \sum_{i} G_i (\sigma_{\psi})>_{N, 
\sigma_{\psi}} .$

As a consistency check, we have reconstructed the APV signal in a 
different way.  $D_{at}$ and ${\cal A}$ measured for
each pulse, yield the probe polarization tilt,  $\theta =
-D_{at}/2(\exp{(\eta {\cal A}}) - 1)$.
for $D_{at}$.
Replacing $D_{at}$ by $\theta$ in Eq.~2 provides a second 
determination of the linear
dichroism {\it G}.
Both methods lead to practically identical final results.

The validity of the entire acquisition procedure has been tested by 
programming the reversal of an additional tilt $\theta_0$ of
the input polarization
$\hat \epsilon_{ex}$, {\it willingly
correlated} with $\vec E_l$ reversal.
It is intended to mimic the APV effect by optical means.
More precisely, the tilt and the field are switched in accordance 
with the following pattern,

\noindent $(\theta_0  +
\theta_{cal}, \vec E_l)$$(\theta_0$ - $\theta_{cal}, \vec E_l)$ (-$\theta_0 +
\theta_{cal}, $ -$\vec E_l)$(-$\theta_0 $ -$\theta_{cal}, $ -$\vec E_l)$

\noindent  expected to mimic $\theta^{pv} = \theta_0$. We expect the 
reconstructed signal {\it G} (Eq. 2)
to give  $\theta_0 + \theta^{pv} $.
This test performed with $\theta_0 /
\theta_{cal} = 10^{-1}$, has been successful within a precision of 2 $\%$.
  When $\hat \epsilon_{ex} $ was rotated by increments of 45$^0$, the 
same procedure enabled
us to test the isotropy of the polarimetry method, with SNR $\sim$ 100.
The $\theta_0 $ tilt  creates a  large linear dichroism $\gamma_1$, 
with no optical rotation $\alpha_2 $. Thus,
we also test successfully the discrimination between $\gamma_1 $ and 
$\alpha_2$, based on the para/ortho reversal.
This is a verification of the reconstitution method (Eq. 2), which 
deals correctly with the problem of
a large gain anisotropy.

  This test also shows the importance of a continuous search for a 
possible unwanted $\vec
E_{l,odd}$ tilt of the input excitation polarization during APV 
measurements. This has been implemented using a
visible polarimeter of the same design as the probe polarimeter, with 
sensitivity about 2.5 times better. Up
to now, a stray effect of this kind, which might have arisen from 
electromagnetic interferences associated with pulsed $\vec E_l$ field
monitoring, remains buried in noise ($\leq  3.5 \times 10^{-8}  \; $rd).

There is one more harmful effect which arises from a longitudinal 
magnetic field reversing with the $\vec E_l$-field,
$B_z(E_{l,odd})$. This causes a Hanle precession ($ E_{l,odd}$) of 
the axes of the atomic alignment
responsible for the parity conserving linear dichroism, around the 
beam axis.  In fact, such a $B_z$ field also gives rise to an optical
rotation by Faraday effect.
The $B_z(E_{l, odd}$) Faraday effect can thus be isolated and exploited
to measure the field magnitude
\cite{bou95}. For this measurement, we select the highly $B_z$-sensitive
$7S_{1/2, F=4} - 6P_{3/2, F=5}$ line. Owing to the progress which has 
resulted from grooving the internal surface of the
alumina cell \cite{gue02}, the $B_z(E_{l,odd})$ field has remained 
either practically absent or small ($ <50 \mu$G) during
the measurements reported here and more importantly it has exhibited 
only very slow drifts. Thus it is
possible to correct for its effect, with only
slight increase of the statistical noise, by inserting field 
measurements between longer periods of APV data averaging.
We have verified the exactitude of the correction by performing 
measurements on both transitions, while willingly applying a
$B_z(E_{l,odd})$ field of known magnitude.

We have calibrated the magnitude of the $E_l$ field inside the 
alumina cell by exploiting the atomic signals\cite{note}. Using this 
calibration
leading to  $E_l$ = 1.619 kV/cm, with 2$\%$ accuracy, and the result 
given by the Boulder
group~\cite{woo97} we expect to obtain for the  $6S_{1/2, F=3} - 
7S_{1/2, F=4}$ excitation line:
$\theta^{pv} =  - \frac{\rm{Im E}_1^{pv}}{\beta E_l} = 0.962\pm 0.020 
\times 10^{-6} $ rd.

We have now accumulated 3200 experimental isotropic values of the 
$E_{l,odd}$ linear dichroism. Fig.~2 shows the results of
successive runs. The final result is:
\begin{equation}
\theta^{pv}_{exp}(\mu\rm{rd}) = 1.082\pm 0.091 ~(stat)   ,
\end{equation}
  which agrees with the expected value within statistical error. The 
chiral optical gain which manifests APV is thus clearly detected.
  This validates our method.
\begin{figure}
\centerline{\epsfxsize=80mm \vspace{-0.3cm} \epsfbox{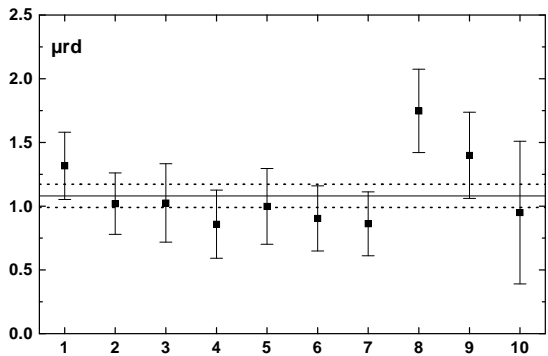}}
\vspace{-0.0cm}
\caption{\footnotesize Experimental values of $\theta 
^{pv}_{exp}$($\mu$rd) obtained in successive runs with
experimental conditions kept as constant as possible. The solid (and 
dashed) lines represent the mean (and the statistical
error on this mean). }
\end{figure}
The experiment has not yet
reached its ultimate sensitivity. The noise is about 2.5 times the shot
noise limit. Its sensitivity to optical adjustments proves that extra 
noise arises from interferences inside the inhomogeneously thick 
sapphire
plates mounted as windows.
In our next Cs cell,
their parallelism will be much improved. By temperature tuning 
\cite{jah01}, an excellent reflexion minimum $\leq 10^{-3}$  can be
achieved and should be accompanied by strong reduction of the 
interference noise.
For further improvement of the SNR, magnification of polarization 
tilts using a dichroic optical component \cite{cha97}, is also a
promising technique we plan to implement soon. In parallel, we are 
constructing a new type of cell with internal ring electrodes
spaced by ceramic braces. This geometry inhibits SEE while the 
electrodes connected at fixed
potentials tend to  evacuate the electrons circulating inside the 
vapor. Hence, we expect a
reduction of the transverse fields~\cite{gue02}, still present during 
the work reported here.
All in all combined with an increased acquisition time (now 210 h) by 
a factor of 3 or 4, these improvements should bring us
to within reach of our
$1\%$ precision objective.

  We thank many colleagues, more particularly F. Lalo\" e, D.~Treille, 
C.~Bouchiat, M. D.~Plimmer, for their interest in this work
and encouragements, L. Pottier for early contributions and A. 
Clouqueur for technical assistance. A.W. acknowledges support from 
CNRS
(IN2P3) and S.S. from the European Commission. Laboratoire Kastler 
Brossel is a Unit\'e de Recherche de l'Ecole Normale Sup\'erieure et 
de
l'Universit\'e Pierre et Marie Curie, associ\'ee au CNRS (UMR 8552).

$^a$ {\small  Present address: Laboratoire de Physique des Lasers
Univ.  Rennes 1, Campus de Beaulieu,
35700 Rennes, France.}

$^b$ {\small  Also at E. Fermi Physics Department, Pisa Univ., Italy.}


\begin{thebibliography}{50}
\bibitem{note}
  This measurement is performed using a circularly polarized 
excitation beam with the  polarimeter operating as a
circular analyzer. Then,  $D^{amp}\propto ((M_1+M_1^{hf})^2 + \beta^2 
E_l^2)$, is
detected with $E_l$ on and off. Since $M_1, M_1^{hf}$ and $\beta $ 
are precisely known~\cite{ben99}, we thus deduce
the magnitude of $E_l$.
\end{thebibliography}
  \end{document}